\documentclass[aps,twocolumn,floatfix,titlepage,balancelastpage,raggedbottom,amsfonts,amssymb,amsmath,showkeys]{revtex4}
\usepackage{epsfig}
\usepackage{bm}
\usepackage{times}
\usepackage{mathptmx}
\usepackage{amsmath}
\usepackage{graphicx}
\usepackage{graphics}
\usepackage{amssymb}
\usepackage{bm}
\usepackage[english]{babel}
\begin{document}
%=======================================================================
\title{Molecular dynamics simulations of the contact angle between water
droplets and graphite surfaces}
\author{Danilo Sergi}
\affiliation{University of Applied Sciences (SUPSI),
The iCIMSI Research Institute,
Galleria 2, CH-6928 Manno, Switzerland}
\author{Giulio Scocchi}
\affiliation{University of Applied Sciences (SUPSI),
The iCIMSI Research Institute,
Galleria 2, CH-6928 Manno, Switzerland}
\author{Alberto Ortona}
\affiliation{University of Applied Sciences (SUPSI),
The iCIMSI Research Institute,
Galleria 2, CH-6928 Manno, Switzerland}

\date{\today}

%================================================
\begin{abstract}
Wetting is a widespread phenomenon, most prominent in a number of cases, both in nature
and technology. Droplets of pure water with initial radius ranging from $20$ to $80$ [\AA]
spreading on graphitic surfaces are studied by molecular dynamics simulations. The equilibrium
contact angle is determined and the transition to the macroscopic limit is discussed using
Young equation in its modified form. While the largest droplets are almost perfectly spherical,
the profiles of the smallest ones are no more properly described by a circle. For the sake of
accuracy, we employ a more general fitting procedure based on local averages.
Furthermore, our results reveal that there is a possible transition to the macroscopic limit.
The modified Young equation is particularly precise for characteristic lengths
(radii and contact-line curvatures) around $40$ [\AA].
\end{abstract}
%=======================================================================
\maketitle
%======================================================================

\section{Introduction}

Wettability is a long-standing issue primarily addressed by
looking at the angle of contact at the edge of the interface between a liquid and a solid
(sessile droplet method) \cite{young}.
In spite of the simplicity of the formulation of the problem, this procedure is at the basis of many
investigations devised to assess the behavior of a liquid or a material
in a number of industrial processes and applications (see
Ref.~\cite{adhesion} for a review), and it has been demanding valuable
efforts, both experimental and theoretical \cite{small,quere,gennes,tau}.
Computer simulations provide useful guidelines for their power to deal with
the complexity of large assemblies of interacting components. In classical
molecular dynamics studies, the systems are described at the molecular
level according to the laws of classical mechanics and electrodynamics.
For these reasons, this approach proves to be both versatile and accurate.
Recent breakthroughs have triggered a burst of interest in the properties of graphene
\cite{nobel1,nobel2,nobel3}.  Promising applications in a variety of fields could indeed
develop. Our interest in graphitic materials is related to their optimal dispersion in
polymeric matrices for the best manufacturing of composite materials \cite{nano_surf,dispersion_surf,review_composite}.
In that respect, the wetting properties of graphitic surfaces by water are of course
the starting point of any subsequent investigation. In particular, the transition to the
macroscopic limit is essential for any further study adopting more advanced modelization schemes
\cite{shinoda1,martini1}.  Our analysis of profiles differs from the
well-established method \cite{binning,werder} in that we approximate them by a more
general curve than a circle. This way of proceeding is especially necessary for small
droplets, exhibiting the largest deviations from the predictions of Young equation
and from a spherical cap. This last aspect has already been recognized
in previous studies \cite{danmark}. Yet, we also propose an analytic expression for the
oxygen-oxygen radial distribution function of water.
%======================================================================

\section{Simulations}

All molecular dynamics simulations are performed
with LAMMPS \cite{pppm}, a code that supports parallelization optimally
\cite{parallel}. Numerical integration is accomplished with the algorithm rRESPA,
allowing to deal with multiple time step sizes \cite{pppm}. We choose a time step
of $2$ [fs] for non-bonded interactions and of $1$ [fs] for bonded
interactions. All Lennard-Jones forces are evaluated with a potential of
the type 12-6. The in-built CHARMM force field \cite{charmm} is employed
to prevent van der Waals interactions from decaying abruptly at the cutoff
distance: the forces are smoothly corrected to zero from $10$ to $12$ [\AA]. 
Pairwise Coulomb interactions within a distance of $12$ [\AA] are calculated in the
real space and beyond this value with a particle-particle, particle-mesh
method; the precision is set to $10^{-4}$. The pairwise interactions among
atoms separated by one or more bonds are neglected. The neighbor lists are
always updated at every time step. These general settings are always applied,
unless specified otherwise.
%======================================================================

\subsection{Water}

Throughout our work we use for the water the SPC/Fw model introduced in
Ref.~\cite{flexible}. We address the reader to this detailed study for the definitions
(partial charges, equilibrium distances, interaction parameters, etc.).
We start from $512$ molecules of water arranged regularly in a cubic
box of side length $30$ [\AA] with periodic boundary conditions.
We let the system evolve for $200$ [ps] in the ensemble NPT
(Nos\'e-Hoover integration). This simulation is performed as equilibration
with a single time step size of $1$ [fs].
The target temperature and pressure are $298.16$ [K] and $1$ [atm],
respectively. Since the main purpose is to reproduce experimental
densities, we choose the parameters that control the convergence so
as to fix the temperature and the volume, while we still let the
pressure fluctuate. It is in fact well-known that the pressure is
a very sensitive function of the volume and difficult to equilibrate
accurately \cite{manual}. We then let the system evolve for other
$0.5$ [ns] at NVE conditions; the final configuration of this dynamics is replicated
and used to obtain the droplets of water.
%======================================================================

\subsection{Droplets and graphitic substrate}

All starting configurations are formed by two parallel planes of
graphene and a semisphere of molecules of water. The planes of
graphene are separated by $3.4$ [\AA] with the lower one translated
by the vector $(l/2,\sqrt{3}l/2)$ with respect to the first; $l=1.42$
[\AA] is the length of the bonds among the carbon atoms. The two
planes are approximately squared with side of at least $30$ [\AA]
larger than the diameter of the semisphere. The semisphere is centered
above the planes of graphene at a distance of $3$ [\AA] from the upper
plane. The boundaries are periodic and two images of the droplets
are separated by at least $100$ [\AA] in the $z$ direction.
The mass of the carbon atoms is $m_{\textrm{C}}=12.011$ [g/mol].
The water and the graphene interact via van der Waals forces
between the atoms of carbon and oxygen with force field parameters
$\epsilon_{\textrm{CO}}=0.0478$ [kcal/mol] and $\sigma_{\textrm{CO}}=3.581$
[\AA] \cite{wang,epsilon}. The system is
equilibrated for $0.5$ [ns] in the ensemble NVT (Nos\'e-Hoover thermostat)
with the temperature of the water maintained at $298.16$ [K]. The system is
studied during a further evolution of $1$ [ns] in the microcanonical
ensemble by gathering frames at every $0.5$ [ps].
%======================================================================

\section{Analysis}

\subsection{Water}

The density of water is calculated using the formula
$\rho=N(m_{\textrm{O}}+2m_{\textrm{H}})/(3\cdot0.602\cdot V_{\mathrm{dom}})$.
$N$ is the total number of atoms and $V_{\mathrm{dom}}$ is the volume in \AA$^{3}$ of the
cubic domain resulting from the preliminary simulation of equilibration. By using
this formula the result is expressed in g/cm$^{3}$.
The oxygen-oxygen radial distribution function $g(r)$ is defined by
$(N/3V_{\mathrm{dom}})g(r)4\pi r^{2}\Delta r=W(r)$.
$W(r)$ is the average number of oxygen atoms in a shell of width
$\Delta r$ at a distance $r$ from a given oxygen atom. We choose
$\Delta r=0.05$ [\AA] and of course $g(r)$ is unitless.
We also calculate the cumulative probability
$P(r)=(3/N)\int^{r}_{0}W(r)n\textrm{d}r$,
which yields the fraction of oxygen atoms within a distance $r$
from a given atom of oxygen. $n=1/\Delta r$ is the number of shells
per unit length. (By writing the above integral as a discrete sum, that
sum would run over the number of shells.)
%======================================================================

\subsection{Droplets}

The contact angle $\theta$ is defined by the tangent at
the contact line, the edge of the interface between the solid and
liquid phases (see Fig.~\ref{fig:profile}). The contact angle of a macroscopic
droplet with spherical symmetry is well described by the equation
$\cos\theta=(\gamma_{\textrm{sv}}-\gamma_{\textrm{sl}})/\gamma_{\textrm{lv}}$.
The $\gamma$'s are the surface/interfacial tensions. The subscripts s, l and v 
stand for solid, liquid and vapor, respectively. The above relation is generally
referred to as Young equation \cite{young}.
Especially for small droplets,
Young equation needs a corrective term that accounts for the line
tension, leading to
$\cos\theta=\cos\theta_{\infty}-\kappa/(\gamma_{\textrm{lv}}R)$.
As the notation suggests,
$\cos\theta_{\infty}=(\gamma_{\textrm{sv}}-\gamma_{\textrm{sl}})/\gamma_{\textrm{lv}}$
is Young equation, which yields the contact angle $\theta_{\infty}$
in the macroscopic limit. $R$ is the radius (or curvature) of the contact
line and $\kappa$ is the line tension. If $\theta<90^{\circ}$, we speak about
hydrophilic behavior; hydrophobic if $\theta>90^{\circ}$. Complete wetting
corresponds to $\theta=0^{\circ}$. The discussion of the contact angle is much
richer than what reported here. A more detailed treatment can be found in
Refs.~\cite{adhesion,small,quere,gennes}.
In order to obtain the profile of the droplets, it is applied the method
explained in Refs.~\cite{binning,werder,epsilon} with $\Delta V=1.9\times 10^{-4}$
[\AA$^{3}$]. Typically, the contact angle is extracted by  superimposing
a circle on the profile coming out from the simulations. The center
and the radius of the circle are obtained by a fit. Here we have
decided to proceed in a quite different way. Indeed, there appears
that the circle departs from the shape of the droplets
in particular for the smallest ones where the contact angle is calculated,
because its radius of curvature is weaker. 
We thus approximate the profile of the droplets by a piecewise linear function. 
The idea is to subdivide the profile into small elements. For the points falling
within an element, we calculate the average values for both $x$ and $y$ coordinates.
These average values are the usual parameters used in linear regressions.
The contact angle is calculated
from the slope of the linear function obtained by a linear regression
on the average points above the contact line at most $5$ [\AA].
The contact line point is assumed to be at $y=2^{1/6}\sigma_{\mathrm{CO}}$.
The contact area is simply given by $C=\pi R^{2}$; the contact line is $L=2\pi R$.
The overall interfacial surface of the droplets is calculated by means of the formula
$S=2\pi r^{2}+2\pi r(r+\sqrt{r^{2}-R^{2}})$ and their volume with
$V=(2/3)\pi r^{3}+\pi r^{2}\sqrt{r^{2}-R^{2}}-(\pi/3)(r^{2}-R^{2})^{3/2}$.
%======================================================================
%=======================================
\begin{table}[t]
\begin{ruledtabular}
\begin{tabular}{cccc}
$L_{\mathrm{dom}}$ [\AA] & $V_{\mathrm{dom}}$ [\AA$^{3}$] & $\rho$ [g/\AA$^{3}$] & $d$ [molecules/\AA$^{3}$]\\
\hline
$24.835$ & $15'318$  & $1.0002$ & $0.0334$
\end{tabular}
\end{ruledtabular}
\caption{
\label{tab:rho}
Length of the side of the cubic simulation domain resulting from equilibration,
its volume, mass and molecular densities.}
\end{table}
%=======================================
\begin{figure}[t]
\includegraphics[width=8.5cm]{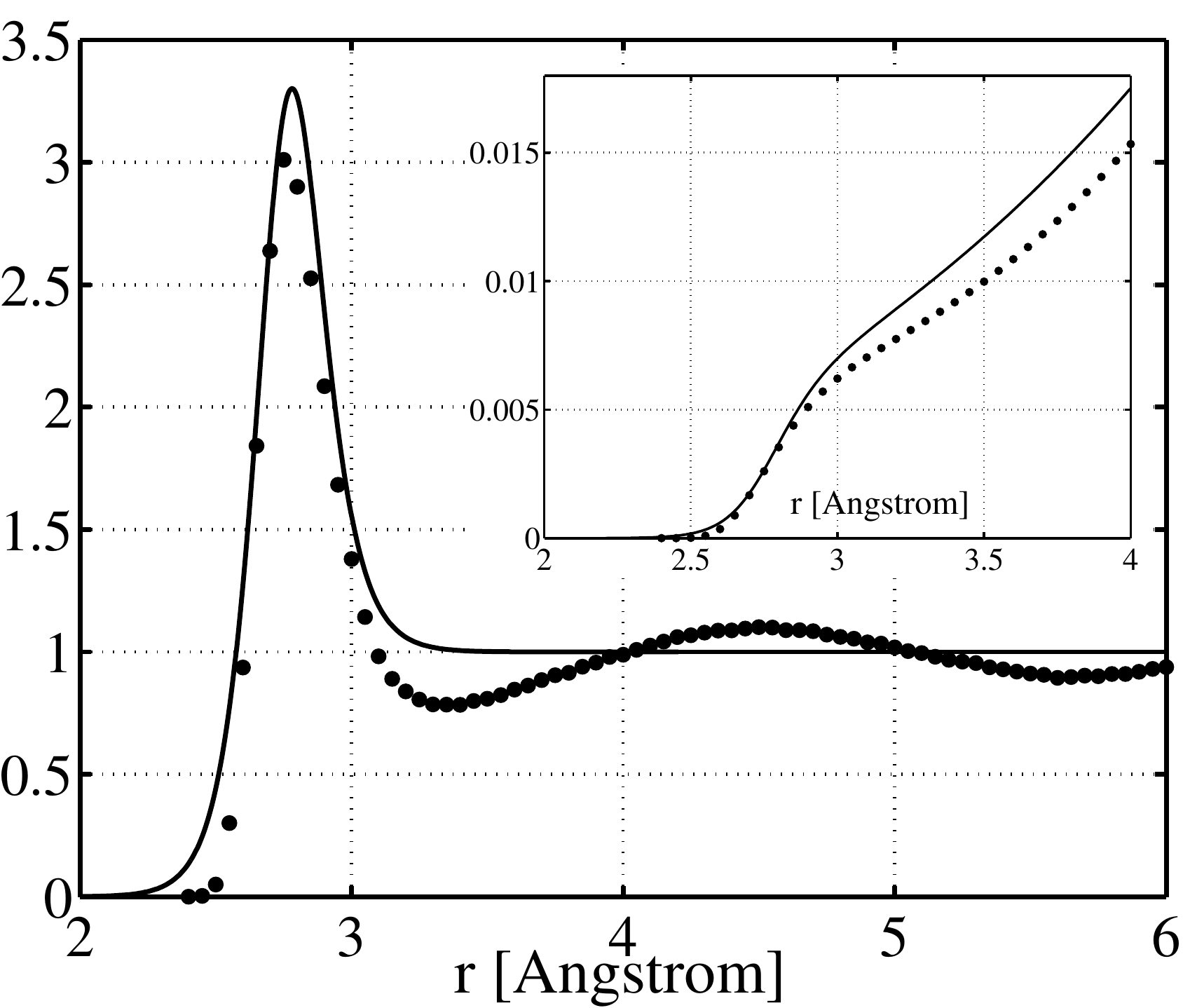}
\caption{\label{fig:dirac}
Oxygen-oxygen radial distribution function of water.
The solid line is the plot of the function $u(r)$, Eq.~\ref{eq:u}.
The parameters $A$ and $B$ are the result of a fit to the data,
represented as filled circles. We find
$A=2.7610\pm 0.0329$ [\AA] and $B=0.0833\pm 0.0183$ [\AA].
Inset: Plot of the function $f(r)$ (see main text), with the same
parameters $A$ and $B$ (solid line) compared to the same data of
Fig.~\ref{fig:p} (filled circles). The two curves differ
at most of $2.25\cdot 10^{-3}$.}
\end{figure}
%=======================================
%=======================================
\begin{figure}[t]
\includegraphics[width=8.5cm]{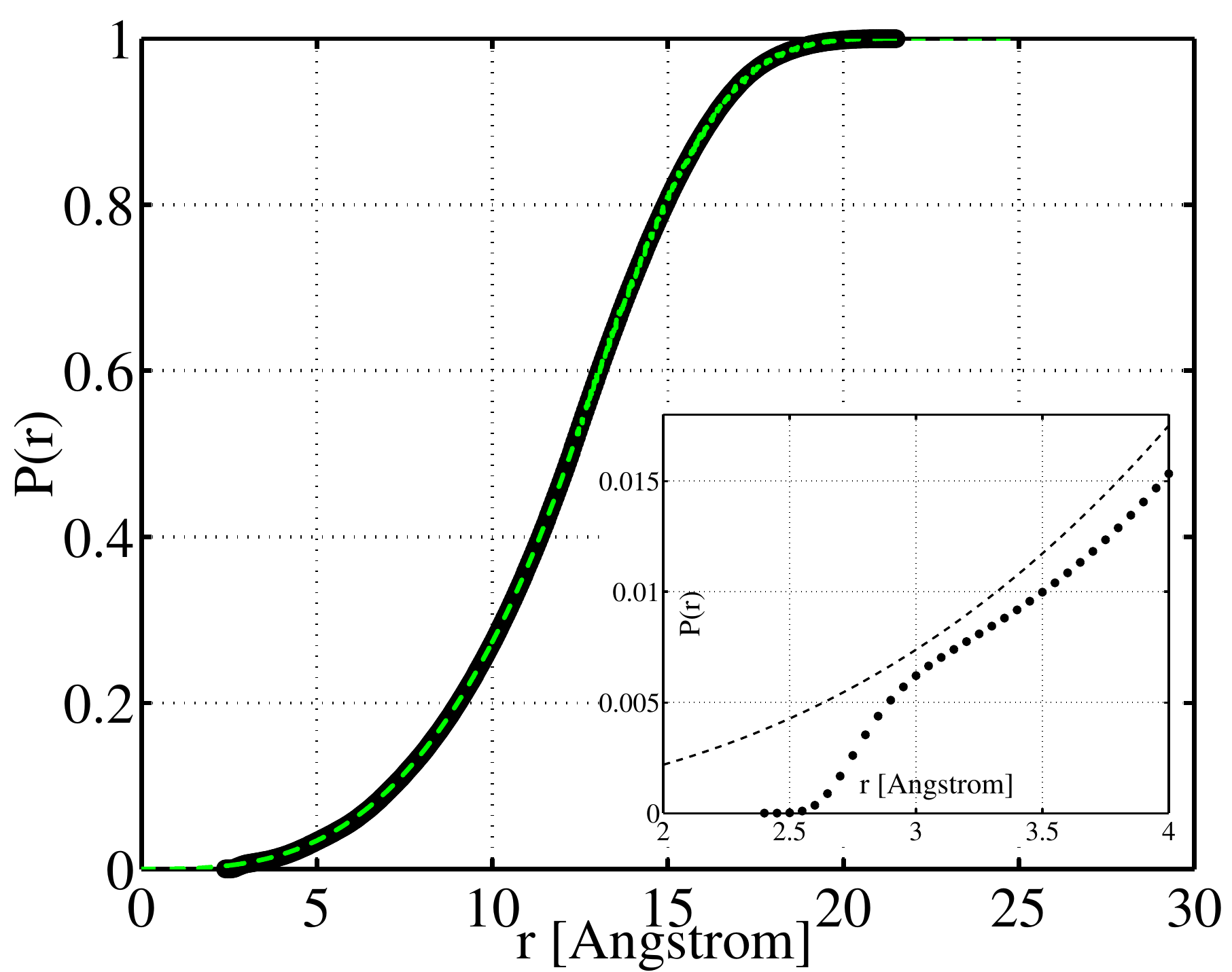}
\caption{\label{fig:p}
Cumulative probability $P(r)$ (see main text).
The dashed line accounts for the continuum, homogeneous behavior.
We start from a sphere centered at the origin. When the radius of the
sphere is larger than $L_{\mathrm{dom}}/2$ (see Tab.~\ref{tab:rho}), the sphere overlaps
with itself because of periodic boundary conditions and we can no more
use the formula $P(r)=(4/3)\pi r^{3}/L_{\mathrm{dom}}^{3}$. We thus employ the Monte Carlo
principle in its elementary form. For a given radius larger than $L_{\mathrm{dom}}/2$,
$30'000$ points are randomly (uniform distribution) placed in a cube
of side $L_{\mathrm{dom}}$ and we count the fraction of them falling within the
portions of the sphere. Inset: Magnification around the position of the
first peak of the radial distribution function (cf.~Fig.~\ref{fig:dirac}).}
\end{figure}
%=======================================

\section{Results and discussion}

\subsection{Water}

Table \ref{tab:rho} summarizes some final results for
the box of SPC/Fw water that is used as source in order to extract the droplets.
Our findings are in good agreement with the original work for that atomistic water model
\cite{flexible}. Figure \ref{fig:dirac} shows the oxygen-oxygen radial distribution function.
The results for the cumulative probability $P(r)$ in Fig.~\ref{fig:p} indicate that
the behavior of water differs slightly from that of a continuum, homogeneous medium,
except in the closest neighborhood of the oxygen atoms. The inset tells us that
technically the first peak
of the radial distribution function is related to the derivative of a step
function (or Heaviside function). We thus consider a function of the type
$f(r)=(4/3)(\pi r^{3}/L_{\mathrm{dom}}^{3})/(1+\textrm{e}^{(A-r)/B})$.
Of course, this function is no more normalized to unity in the interval $[0,L_{\mathrm{dom}})$,
but it provides a good approximation of the cumulative distribution $P(r)$
where the first peak of the radial distribution function occurs. The
denominator of $f(r)$ is reminiscent of the Fermi-Dirac distribution
function, which is a step function at low temperatures. The
parameter $A$ fixes the position of the first peak and the parameter $B$
determines its width and height. After a simple calculation, we find for the
radial distribution function the following analytic expression:
\begin{equation}
u(r)=\frac{1}{1+\textrm{e}^{(A-r)/B}}\Big[1+\frac{r}{3B}
\frac{\textrm{e}^{(A-r)/B}}{1+\textrm{e}^{(A-r)/B}}\Big]\ .
\label{eq:u}
\end{equation}
This function reproduces correctly the first peak, as shown in
Fig.~\ref{fig:dirac}. The main discrepancy with the data for
the SPC/Fw model is in the neighborhood of the first minimum, corresponding
to a density depletion with respect to the bulk value. The above method is of course
unable to explain this aspect of the fine structure of water.

%===========================================================================
%=======================================
\begin{figure}[t]
\includegraphics[height=8.5cm]{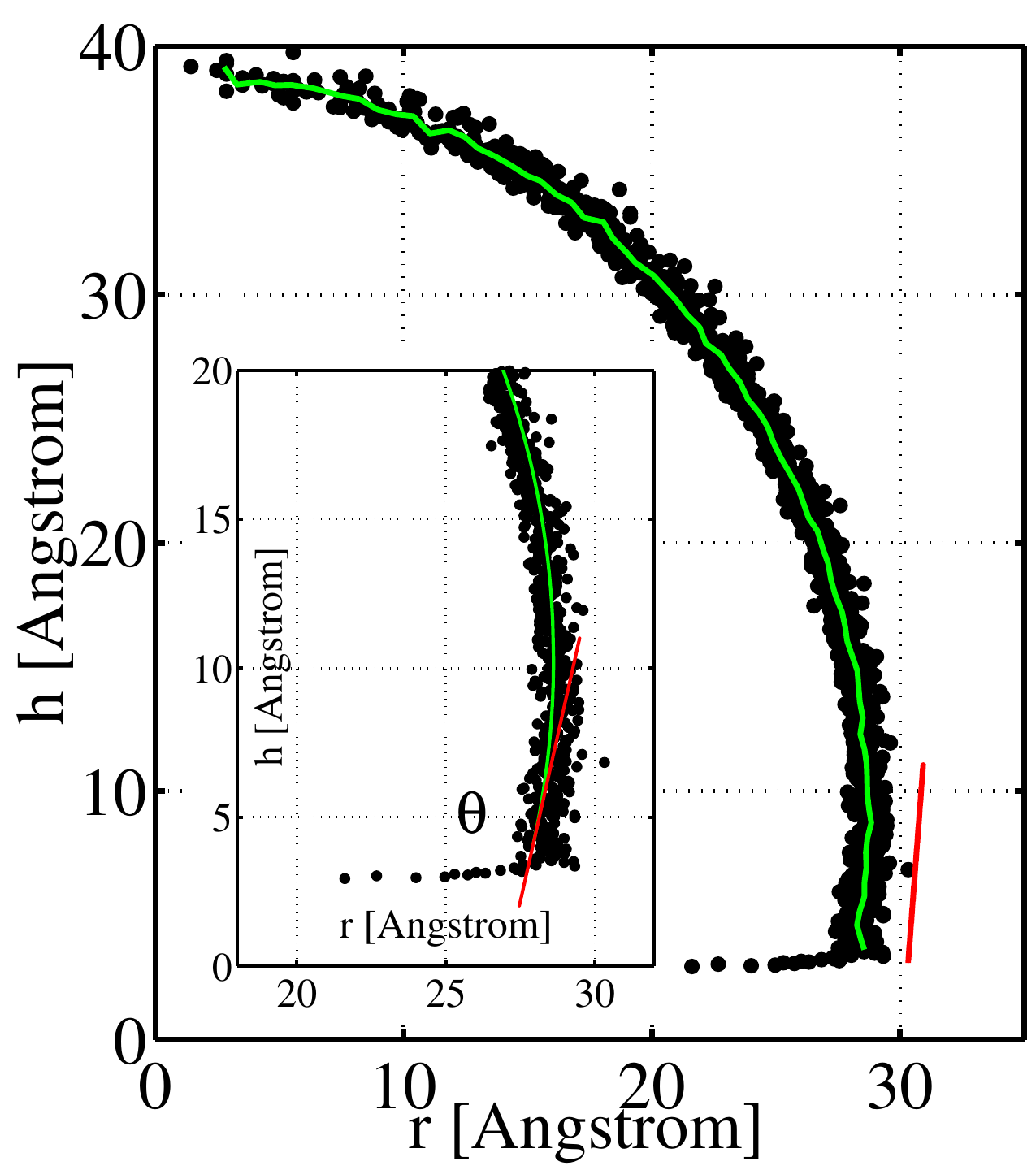}
\caption{\label{fig:profile}
Profile of the droplet of initial radius of $30$ [\AA].
The green curve represents the profile of the droplet
as obtained from local averages.
The straight line is a guide for the eyes and its slope
amounts to the value used for determining the contact angle. Inset: Profile from
a circular fit to the data in the neighborhood of contact line. The straight line is
the tangent at the contact-line point.}
\end{figure}
%=======================================
%=======================================
\begin{figure}[t]
\includegraphics[width=8.5cm]{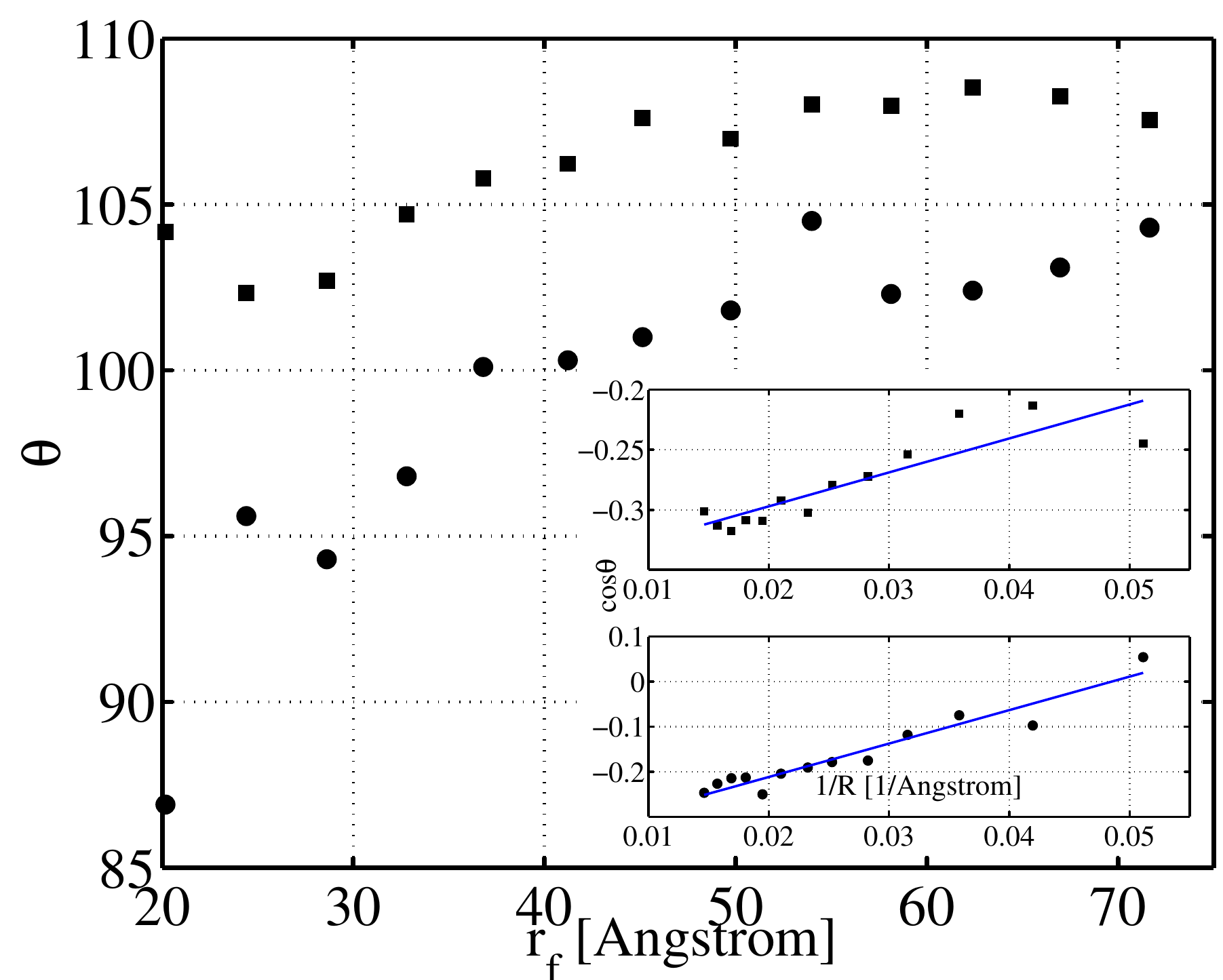}
\caption{\label{fig:theta2}
Contact angle dependence on the final radius from the two approximations of
the droplet profiles (squares for circular fits and circles for local averages).
Inset: Fitting of the data to modified Young equation with contact angles resulting from
circular fits, Top, and from local averages, Bottom.}
\end{figure}
%=======================================
%=======================================
\begin{figure}[t]
\includegraphics[width=8.5cm]{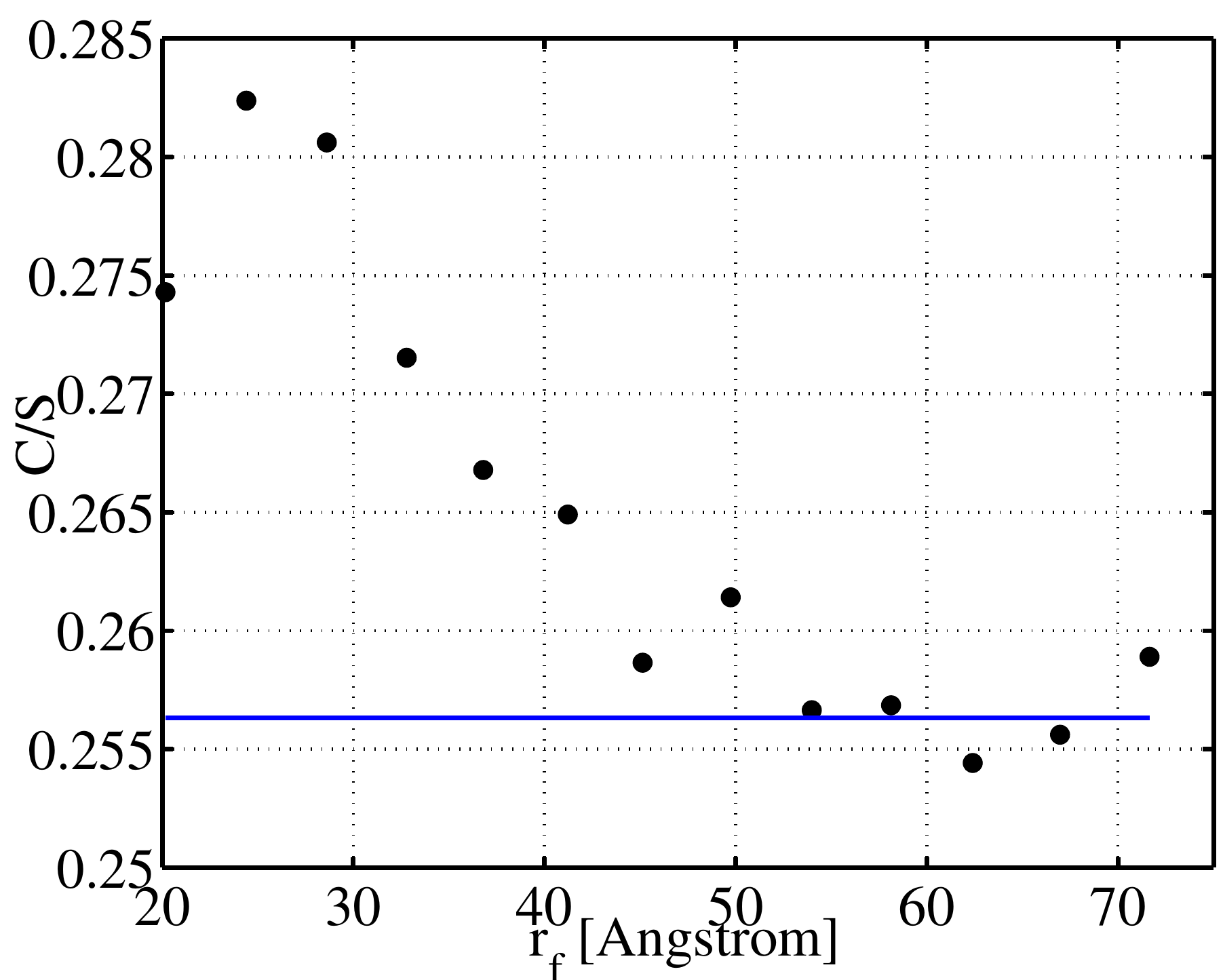}
\caption{\label{fig:ratio}
Contact area to overall interfacial surface ratio as a function of
the final radius of the droplets. The straight line is the macroscopic expectation
under the assumption of a perfectly spherical profile. }
\end{figure}
%=======================================
%=======================================
\begin{figure*}[t]
\includegraphics[width=5.2cm]{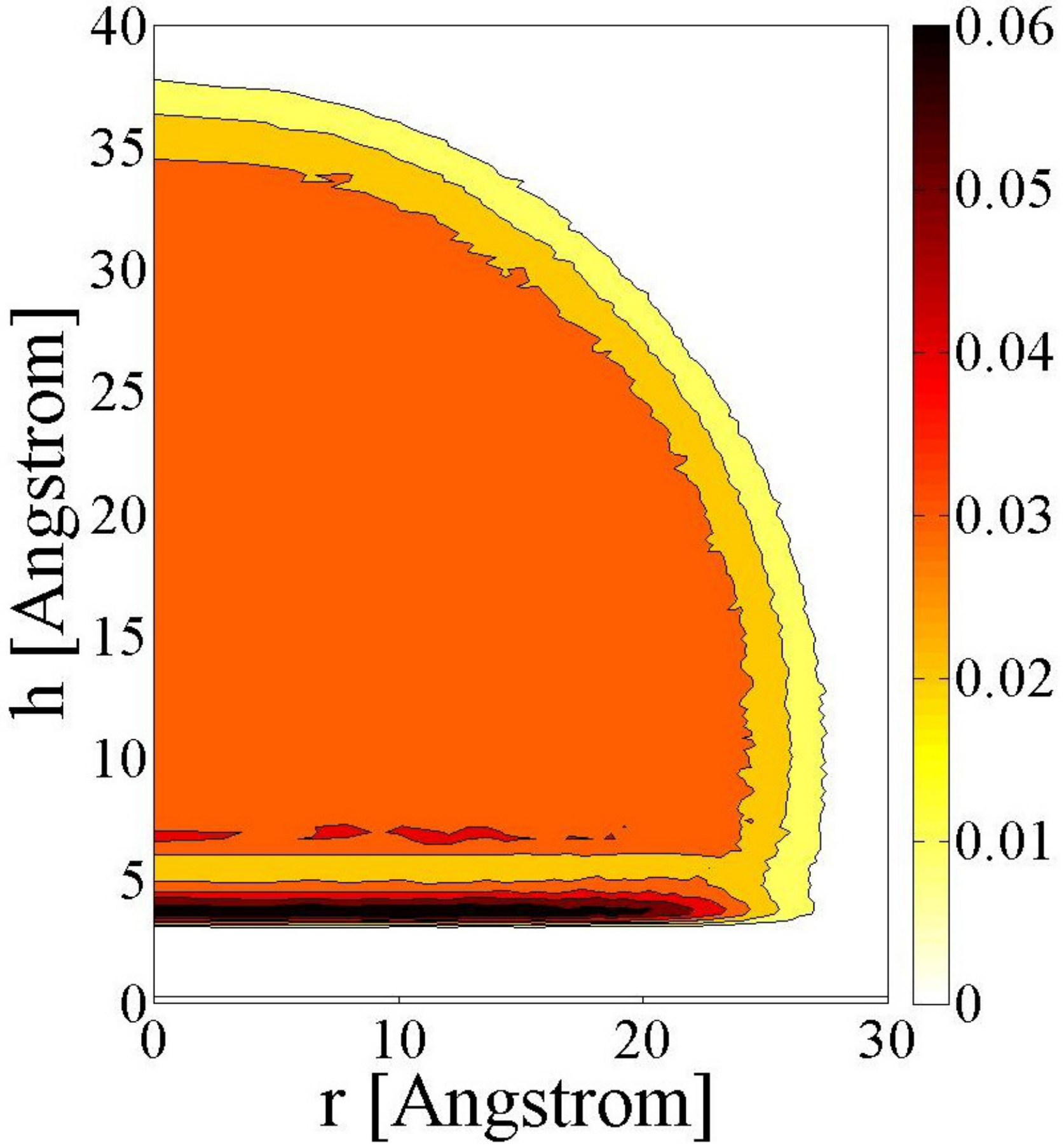}\hspace{0.5cm}
\includegraphics[width=5.2cm]{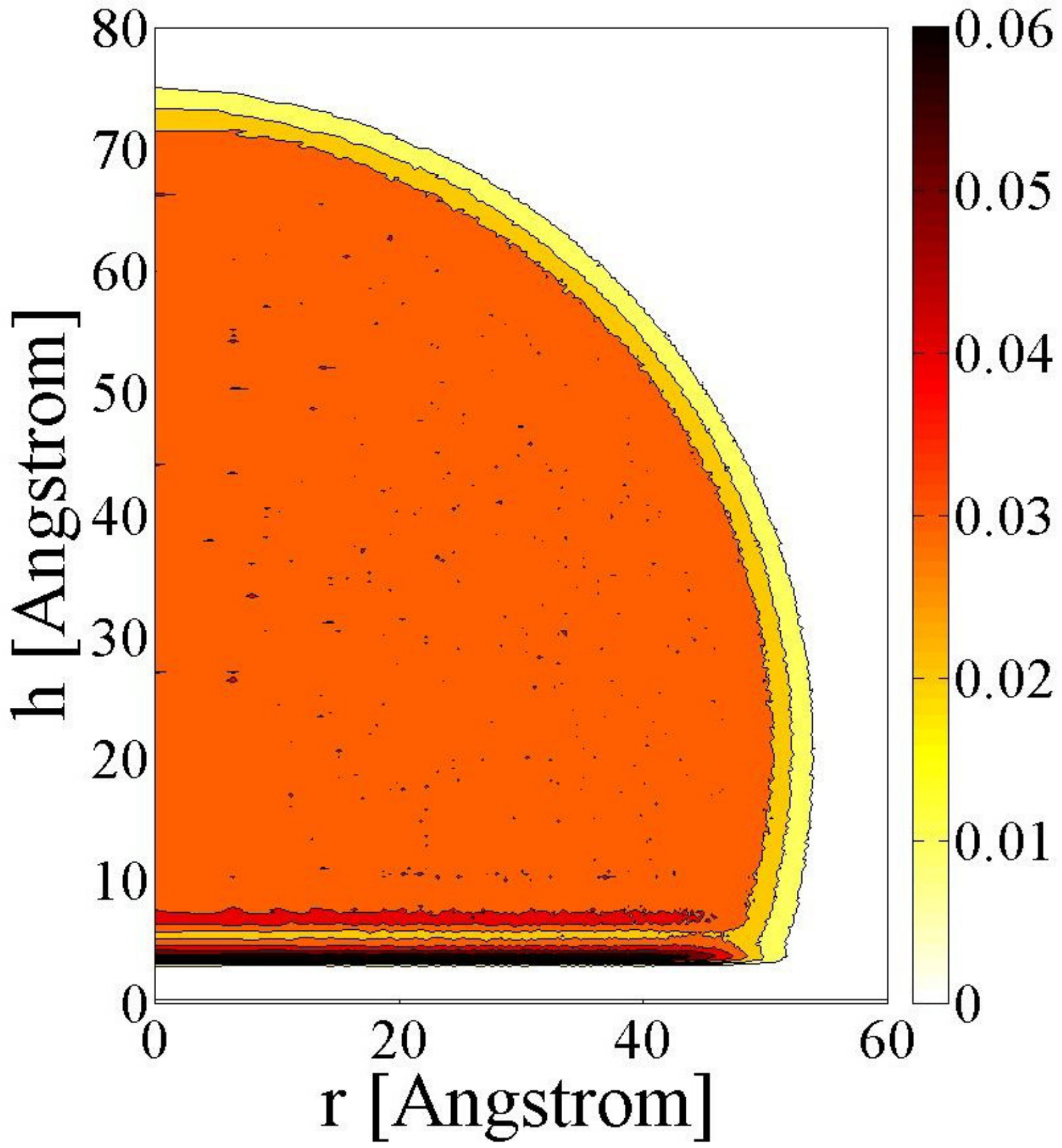}\hspace{0.5cm}
\includegraphics[width=5.2cm]{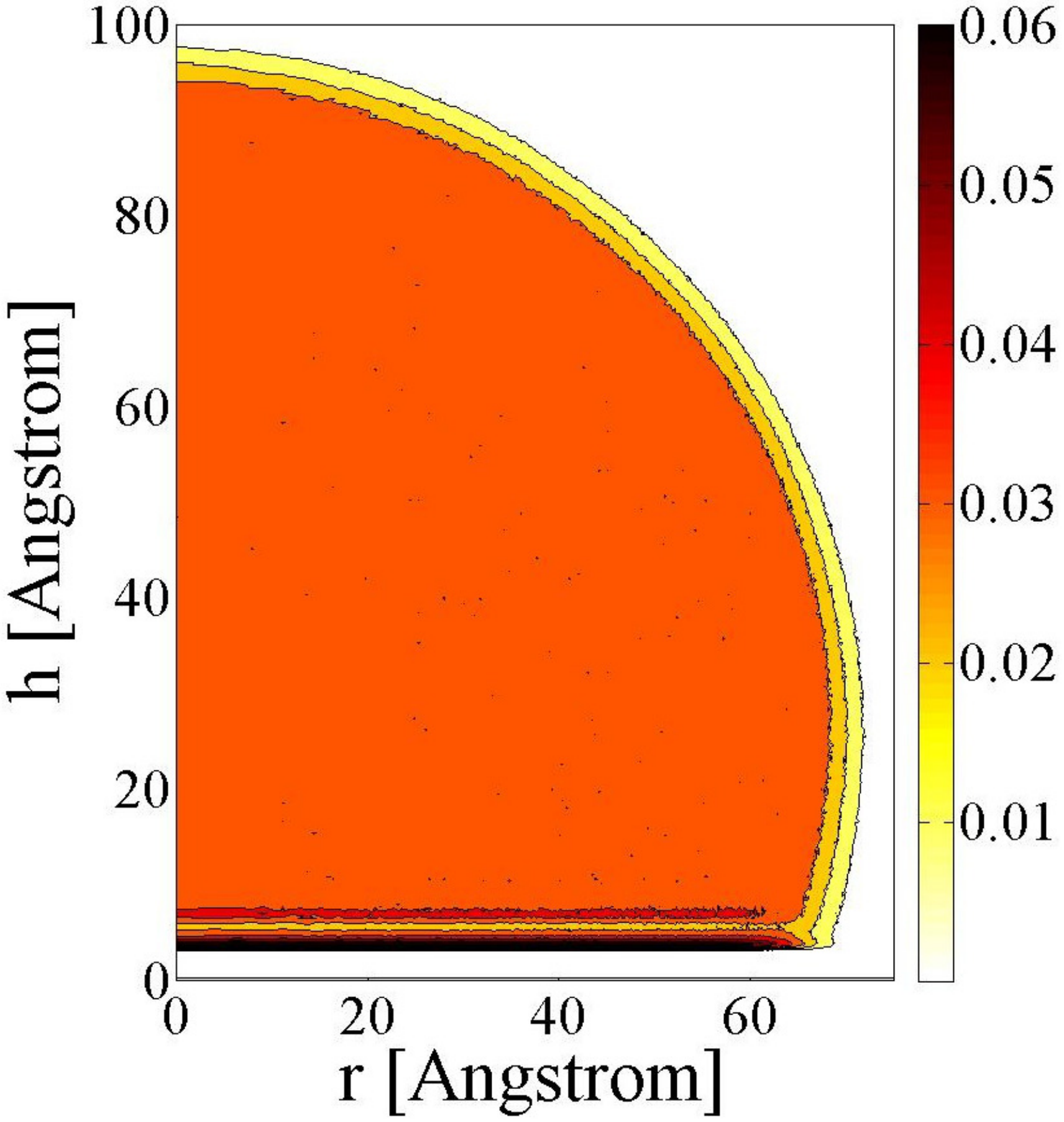}
\caption{\label{fig:density}
Density maps for the droplets of initial radii $30$, $60$ and $80$ [\AA], in this order.
Color code based on molecular density. The thickness of the interface is in all 
cases around $3$ [\AA]. Concerning the first map, the horizontal shift relative to the 
representation of Fig.~\ref{fig:profile} is due to the fact that the first radial bin 
coincides with zero and the coarser binning necessary for this kind of representation.}
\end{figure*}
%=======================================

\subsection{Droplets}

Figure \ref{fig:profile} shows the profile of a small
droplet when approximated by a piecewise linear function. Near the contact line, it
appears that this procedure is more precise than a circular fit. From local averages,
the resulting contact angle is always lower (see Fig.~\ref{fig:theta2}). In both cases, the macroscopic contact angle can
be extracted from an average over the largest droplets, those of radii $60-80$ [\AA], leading to $\theta_{\infty}=103.1^{\circ}$
($\theta_{\infty}=108.1^{\circ}$ for circular fits). Its standard deviation is $0.9\%$ ($0.3\%$) of this average value. 
The residual difference between the two methods indicates that, even for the
largest droplets, in the close neighborhood of the contact line, the profile still 
deviates from a perfectly spherical cap. It turns out that the method based on circular fits
underestimates the base radius. The average bulk molecular density is
$0.0323$ [molecules/\AA$^{3}$] (cf.~Tab.~\ref{tab:rho}), with standard deviation $3.5\%$
of it. By fitting the data to the modified Young equation it is extrapolated a value of $\theta_{\infty}=111.1^{\circ}$
($110.7^{\circ}$ for circular fits).
Especially for the circular profiles, we prefer the first method because the results indicate that around
$60$ [\AA] occurs a transition to the macroscopic limit (see Inset of Fig.~\ref{fig:theta2} and previous basic statistics).
From a typical value for the surface tension of water of $\gamma=72$ [mN/m], for the line tension
it is found $\kappa=-2.04\cdot 10^{-11}$ [N] from circular fits and $\kappa=-5.35\cdot 10^{-11}$ [N] from local averages.
This means that the forces tending to expand the contact area are higher according to the method based on local averages.
In Fig.~\ref{fig:ratio} we compare the contact area to interfacial area ratio, i.e.~$C/S$, with the results
for the droplets having the same final radius but with contact angle $\theta_{\infty}=108.1^{\circ}$
(the value of $110.7^{\circ}$ underestimates significantly $C/S$ for the largest droplets).
Given a final radius, the curvature of the contact line is determined numerically
so as to have the macroscopic contact angle $\theta_{\infty}$. With respect to the
macroscopic expectation, it is found that  for the five smaller droplets on average
the contact area expands of $4.2\%$, the interfacial surface shrinks of $2.8\%$ and
the volume contracts of $7.2\%$. 
Figure \ref{fig:density} compares the density maps of three droplets of different size.
These representations suggest that, for small droplets, a  higher fraction of molecules is involved
in density fluctuations at the interface. A simple calculation shows that this aspect
effectively occurs if $N_{1}/N_{2}>\rho_{1}/\rho_{2}$ holds. The index $1$ designates
a large droplet and $2$ a smaller one. $N$ is the number of atoms in the fluid phase
and $\rho$ the bulk density. If we compare the droplets of initial radii $30$ and $60$ [\AA],
we find $(3\cdot 15'119)/(3\cdot  1'902)>0.0338/0.0337=1$. Since even for the smallest
droplet of initial radius of $20$ [\AA] the density of water is close to the bulk value
and the surface thickness is around $3$ [\AA], we conclude that for sure larger droplets 
have a reduced fraction of molecules at the interface. In other words, as the droplet increases in size, 
its interface grows and the fraction of molecules fluctuating at the interface becomes smaller. The droplet 
of initial radius $60$ [\AA] was also simulated at different temperatures up to $450$ [K]. For this 
temperature increase, it is found that the contact angle varies with good approximation linearly, 
as well as the molecular density. In contrast to what reported in Ref.~\cite{temp} the contact angle 
varies over this temperature range only of a few degrees. Preliminary results using coarse-grained 
models confirm our trend.
%===========================================================================

\section{Conclusions}

Contact angle measurements and comparisons with the predictions of Young equation are generally
carried out under the assumption of a spherical shape of droplets. Figure \ref{fig:ratio} shows
that, when the contribution of the corrective term to Young equation and/or other size effects
\cite{tau,danmark,epsilon} is more important, the droplets deviate significantly from the
macroscopic expectation. Furthermore, below the initial radius of $30$ [\AA], the
contact angle can no more be derived accurately from the tangent to a
circular profile at the contact line (see Figs.~\ref{fig:profile} and \ref{fig:theta2}). 
Actually, from our analysis it clearly emerges that, for small droplets, fluctuations of the surface 
thickness are of major relevance, resulting in the deformation of their spherical shape. For small
droplets the effect is more marked presumably because of their reduced size and the short-range 
nature of non-bonded interactions (cf. Ref. [24]): cohesive forces near the contact 
line are weaker and the contact area would tend to expand, leading to lower contact angles.
Interestingly, it also appears that the macroscopic limit is not reached gradually, but with 
a possible transition around the initial radius of $60$ [\AA]. On the other hand, the predictions of the
modified Young equation are more precisely recovered for initial radii around $40$ [\AA].
Finally, the SPC/Fw model for water is extensively validated \cite{flexible} and the Lennard-Jones parameter
$\varepsilon_{\mathrm{CO}}$ used here resulted from a previous calibration according to recent
measurements carried out under the most ideal conditions \cite{wang,epsilon}. Our
macroscopic contact angle $\theta_{\mathrm{\infty}}$ of $111.1^{\circ}$ (circular fits and extrapolated value)
underestimates the experimental value of $127^{\circ}$ \cite{wang}, taken as reference for calibration \cite{epsilon}. 
We ascribe this discrepancy mostly to the different cutoff scheme and the inclusion of long-range 
interactions \cite{cutoff}. Clearly, our analysis of profiles and the main conclusions drawn for the 
smallest droplets in the hydrophobic regime are robust under small changes of the simulation settings. 
Indeed, the profiles of these droplets will be affected to a larger extent by the fluctuations of the 
surface thickness and in turn no more fully spherical. The accuracy implied by the various simulation 
settings confer a substantial degree of confidence to the findings reported here.
%===========================================================================

\section{Nomenclature}

\begin{tabbing}
\hspace{1cm}\=\hspace{2cm}\=\kill
$A$,$B$\>\AA\>\begin{tabular}{p{5.6cm}}model parameters
\end{tabular}\\
$C$\>\AA$^{2}$\>\begin{tabular}{p{5.6cm}}contact area of droplets
\end{tabular}\\
$d$\>\AA$^{-3}$\>\begin{tabular}{p{5.6cm}}molecular density
\end{tabular}\\
$f$,$u$\>-\>\begin{tabular}{p{5.6cm}}model functions
\end{tabular}\\
$g$\>-\>\begin{tabular}{p{5.6cm}}radial distribution function
\end{tabular}\\
$L$\>\AA\>\begin{tabular}{p{5.6cm}}length of contact line of droplets
\end{tabular}\\
$L_{\mathrm{dom}}$\>\AA\>\begin{tabular}{p{5.6cm}}side length of simulation domain
\end{tabular}\\
$l$\>\AA\>\begin{tabular}{p{5.6cm}}carbon bond length in graphene
\end{tabular}\\
$m$\>g$\cdot$mol$^{-1}$\>\begin{tabular}{p{5.6cm}}mass of atoms
\end{tabular}\\
$N$\>-\>\begin{tabular}{p{5.6cm}}number of atoms
\end{tabular}\\
$n$\>\AA$^{-1}$\>\begin{tabular}{p{5.6cm}}number of spherical shells per unit length
\end{tabular}\\
$P$\>-\>\begin{tabular}{p{5.6cm}}cumulative radial distribution for oxygen atoms
\end{tabular}\\
$R$\>\AA\>\begin{tabular}{p{5.6cm}}base radius of droplets
\end{tabular}\\
$r$\>\AA\>\begin{tabular}{p{5.6cm}}radius of droplets
\end{tabular}\\
$\Delta r$\>\AA\>\begin{tabular}{p{5.6cm}}width of a spherical shell
\end{tabular}\\
$S$\>\AA$^{2}$\>\begin{tabular}{p{5.6cm}}overall interfacial surface of droplets
\end{tabular}\\
$V$\>\AA$^{3}$\>\begin{tabular}{p{5.6cm}}volume of droplets
\end{tabular}\\
$V_{\mathrm{dom}}$\>\AA$^{3}$\>\begin{tabular}{p{5.6cm}}volume of simulation domain
\end{tabular}\\
$W$\>-\>\begin{tabular}{p{5.6cm}}average number of oxygen atoms in a spherical shell
\end{tabular}\\
$x$,$y$,$h$\>\AA\>\begin{tabular}{p{5.6cm}}cartesian coordinates
\end{tabular}\\
$\varepsilon$\>kcal$\cdot$mol$^{-1}$\>\begin{tabular}{p{5.6cm}}interaction parameter for Lennard-Jones potential
\end{tabular}\\
$\gamma$\>N$\cdot$m$^{-1}$\>\begin{tabular}{p{5.6cm}}surface/interfacial tension
\end{tabular}\\
$\kappa$\>N\>\begin{tabular}{p{5.6cm}}line tension
\end{tabular}\\
$\sigma$\>\AA\>\begin{tabular}{p{5.6cm}}interaction parameter for Lennard-Jones potential
\end{tabular}\\
$\rho$\>g$\cdot$cm$^{-3}$\>\begin{tabular}{p{5.6cm}}mass density
\end{tabular}\\
$\theta$\>-\>\begin{tabular}{p{5.6cm}}contact angle
\end{tabular}\\
\end{tabbing}

%===========================================================================
\begin{acknowledgments}
This is work supported by the Swiss Innovation Promotion Agency (KTI/CTI)
under grant P.~No.~10055.1 (BiPCaNP project). Computations were done with the
facilities of CSCS and iCIMSI-SUPSI. We thank their staff for assistance.
We are also grateful to the anonymous Referees for their comments on a previous
version of this work.
\end{acknowledgments}
%===========================================================================

%=======================================================================
\end{document}